\documentclass[12pt]{article}

\newcommand{\blind}{0}

\usepackage[margin=1in]{geometry}
\usepackage{amssymb,amsmath}
\usepackage{graphicx}
\usepackage{natbib}
\usepackage{url}
\usepackage{booktabs}
\usepackage{setspace}
\usepackage{bm}

\usepackage{adjustbox}
\usepackage{array}

\usepackage{multirow}
\usepackage{caption}

\newtheorem{thm}{Theorem}
\newtheorem{dfn}{Definition}

\usepackage{algorithm}
\usepackage[noend]{algpseudocode}

\allowdisplaybreaks

\newcolumntype{R}[2]{%
    >{\adjustbox{angle=#1,lap=\width-(#2)}\bgroup}%
    l%
    <{\egroup}%
}

\pdfoutput=1
\usepackage[breaklinks]{hyperref}
\hypersetup{
    colorlinks=true,
    linkcolor=black,
    citecolor=black,
    filecolor=black,
    urlcolor=black,
}

\usepackage{color}

\usepackage[normalem]{ulem}

\begin{document}

\renewcommand{\baselinestretch}{1}\small\normalsize

\newcommand{\mytitle}{Search Algorithms and Loss Functions for Bayesian Clustering}

\if0\blind
{
  \title{\bf \mytitle}
  \author{
      David B.\ Dahl
      \hspace{0.2cm}\\
    Department of Statistics, Brigham Young University\\
    and \\
    Devin J.\ Johnson\\
    Department of Statistics, Brigham Young University\\
    and \\
    Peter M\"uller\\
    Department of Statistics and Data Sciences, Univeristy of Texas at Austin}
  \date{May 7, 2021}
  \maketitle
} \fi
\if1\blind
{
  \title{\bf \mytitle}
  \maketitle
} \fi

\bigskip
\begin{abstract}
We propose a randomized greedy search algorithm to find a point estimate for a
random partition based on a loss function and posterior Monte Carlo samples.
Given the large size and awkward discrete nature of the search space, the
minimization of the posterior expected loss is challenging.  Our approach is a
stochastic search based on a series of greedy optimizations performed in a
random order and is embarrassingly parallel.  We consider several loss
functions, including Binder loss and variation of information. We note that
criticisms of Binder loss are the result of using equal penalties of
misclassification and we show an efficient means to compute Binder loss with
potentially unequal penalties.  Furthermore, we extend the original variation of
information to allow for unequal penalties and show no increased computational
costs.  We provide a reference implementation of our algorithm.  Using a variety
of examples, we show that our method produces clustering estimates that better
minimize the expected loss and are obtained faster than existing methods.
\end{abstract}

\noindent%
\vfill

\newpage
\renewcommand{\baselinestretch}{1.5}\small\normalsize

\section{Introduction}
\label{sec:intro}

In a typical Bayesian analysis, a great deal of computational effort is spent on
``fitting the model,'' such as sampling from the posterior distribution or
finding a tractable approximation to the posterior distribution.  This, however,
is only part of the inference problem. It is also necessary to summarize the
posterior distribution in order to convey meaningful results. In many problems,
parameters of interest often lie in a subset of $\mathbb{R}^n$ and, depending on
the loss function, the Bayes rule might be the mean or median, which can easily
be derived from posterior samples.  Increasingly, parameters with a more
complicated structure are being considered.  It is often less clear how to
summarize the posterior distribution of these more complicated structures.  In
this paper, we focus on partitions and address the problem of point estimation
from a partition distribution based on samples.

A partition $\rho = \{S_1,\ldots,S_p\}$ of integers $\{1,\ldots,n\}$ is a
collection of subsets (i.e., clusters) such that the subsets are mutually
exclusive, nonempty, and exhaustive. In model construction, partitions are often
used to arrange data $y_1,\ldots,y_n$ such that data within a cluster are
homogeneous. Items $i$ and $j$ are clustered together if $i \in S$ and $j \in S$
for some subset $S \in \rho$. A partition can alternatively be represented by
cluster labels. We say that items $i$ and $j$ belong to the same cluster if and
only if their cluster labels $c_i$ and $c_j$ are equal. We use the terms
clustering and partition interchangeably. Likewise, the terms cluster and subset
are used interchangeably.  As a notational convention, we use cluster labels $1,
\ldots, p$ for the elements of $\bm{c} = (c_1,\ldots,c_n)$ when a partition
$\rho = \{S_1, \ldots, S_p \}$ has $p$ subsets.

Under the Bayesian paradigm, the canonical approach to choosing an estimator is
to introduce a loss function and then report the Bayes rule that minimizes the
posterior expectation of the chosen loss function.  We have:
\begin{equation}
\label{eq_generic_loss}
\hat\rho^* = \underset{\hat\rho}{\text{argmin}}\hspace{3pt}\mathbb{E}(L(\rho, \hat\rho) \mid \mathcal{D})
\quad \quad \text{or} \quad \quad
\hat{\textbf{c}}^* = \underset{\hat{\textbf{c}}}{\text{argmin}}\hspace{3pt}\mathbb{E}(L(\textbf{c}, \hat{\textbf{c}}) \mid \mathcal{D}),
\end{equation}
where $\mathcal{D}$ represents data and $L(\rho, \hat\rho)$ and $L(\textbf{c},
\hat{\textbf{c}})$ are the same loss function represented in the partition and
cluster label notations, respectively. Without loss of generality, we assume
that if the estimator ($\hat\rho$ or $\hat{\bm{c}}$) is equal to the true
parameter ($\rho$ or $\bm{c}$), then the loss function evaluates to zero.
Otherwise, the loss is some positive number representing the economic cost or
regret associated with the decision ($\hat\rho$ or $\hat{\bm{c}}$) in light of
the truth ($\rho$ or $\bm{c}$).

Except in a situation with trivially small sample size $n$, the posterior
expectation in (\ref{eq_generic_loss}) must be approximated, usually using
posterior samples:
\begin{equation}
    \label{eq_posterior_expectation_via_MC}
    \mathbb{E}(L(\rho, \hat\rho) \mid \mathcal{D}) \approx \frac{1}{H} \sum_{h=1}^H L(\rho^{(h)}, \hat\rho)
    \quad \quad \text{or} \quad \quad
    \mathbb{E}(L(\textbf{c}, \hat{\textbf{c}}) \mid \mathcal{D}) \approx \frac{1}{H} \sum_{h=1}^H L(\bm{c}^{(h)}, \hat{\bm{c}}),
\end{equation}
where $\rho^{(1)},\ldots,\rho^{(H)}$ or $\bm{c}^{(1)},\ldots,\bm{c}^{(H)}$ are
$H$ samples from a posterior distribution $p(\rho \mid \mathcal{D})$ or
$p(\bm{c} \mid \mathcal{D}$). These are often obtained from several Markov chain
Monte Carlo (MCMC) chains and may require considerable effort.  Here we merely
assume these are available, and our task is to use these samples to obtain an
estimate that summarizes the partition distribution.

In Section \ref{sec:criteria}, we review the most common partition loss
functions and related criteria, including \citet{binder.1978} loss and the
variation of information \citep{meila.2007, vinh.etal.2010, wade.ghah.2018}.
\citet{wade.ghah.2018} and \citet{rastelli.friel.2018} conclude that Binder loss
\emph{overestimates} the number of clusters, but we note that the applied
literature has almost exclusively used equal costs of misclassification.  We
offer an efficient technique in Section \ref{sec:generalized} to compute Binder
loss with unequal costs of misclassification and show in Section
\ref{sec:verifications} that this addresses the concerns about Binder loss
finding too many clusters. Further in Section \ref{sec:generalized}, we
introduce a novel extension to variation of information (VI) that maintains
theoretical properties of this original loss function, yet allows unequal costs
of misclassification (analogous to the flexibility in Binder loss when using our
efficient technique) and addresses the issue that VI may \emph{underestimate}
the number of clusters.

Selecting a loss function and generating posterior samples allows us to compute
a Monte Carlo estimate of the posterior expected loss, but a far more
challenging practical problem is searching the vast space of partitions for the
minimizer of the Monte Carlo estimate of the posterior expected loss.  We review
the literature of existing search algorithms in Section \ref{sec:algorithms},
and in Section \ref{sec:salso} present SALSO, a new search algorithm for any
partition loss function.  We show in Section \ref{sec:verifications} that SALSO
is substantially faster and leads to demonstrably better estimates than existing
algorithms, yet has tractable complexity such that it readily scales in the
sample size $n$. The new and existing loss functions and the novel
search algorithm are implemented in our \texttt{salso} package on
the Comprehensive R Archive Network (CRAN).

\section{Existing Partition Loss Functions and Other Criteria}
\label{sec:criteria}

Here we describe existing loss functions to estimate a random partition, as well as
related criteria.  Perhaps the simplest loss function is the 0-1 loss function,
defined as $ L_{\text{0-1}}(\rho,\hat{\rho}) = \mathbb{I}\{\rho=\hat\rho\}$ or
$L_{\text{0-1}}(\textbf{c},\hat{\textbf{c}}) =
\mathbb{I}\{\textbf{c}=\hat{\textbf{c}}\},$ where $\mathbb{I}\{A\}$ is the
indicator function equaling $1$ if $A$ is true and $0$ otherwise. The 0-1 loss
function yields the maximum \emph{a posteriori} (MAP) partition, the mode of the
posterior partition distribution. As in many other contexts in Bayesian
analysis, the mode may not well represent the ``center'' of the partition
distribution and the loss functions presented below are generally preferred over
the 0-1 loss function.

\begin{table}[t]
    \caption{Contingency table of counts used to compute various loss functions
    when estimating a population partition $\rho = \{S_1,\ldots,S_p\}$ with an
    estimated partition $\hat\rho = \{T_1,\ldots,T_q\}$.\label{tab:contingency}}
\label{contingency_table} \centering \vspace{2ex}
    \begin{tabular}{@{}cc|cccc|c}
        &\multicolumn{5}{c}{Estimated partition $\hat\rho $} & \\
        \multicolumn{1}{c}{} & & $T_1$ & $T_2$ & $\hdots$ & $T_q$ & \\ \cmidrule{2-7}
        \multirow{5}*{\rotatebox{90}{True partition $\rho$}}
        &$ S_1 $ &  $ n_{11} = |S_1 \cap T_1 | $ & $ n_{12} = |S_1 \cap T_2 | $ & $ \hdots $  & $ n_{1 q} = |S_1 \cap T_q | $ &  $ n_{1\cdot} = | S_1 | $\\
        &$ S_2 $ &  $ n_{21} = |S_2 \cap T_1 | $ & $ n_{22} = |S_2 \cap T_2 | $ & $ \hdots $  & $ n_{1 q} = |S_2 \cap T_q | $ &  $ n_{2\cdot} = | S_2 | $\\
        &\vdots &  \vdots & \vdots &  & \vdots &  \vdots \\
        &$ S_p $ &  $ n_{p1} = | S_p \cap T_1 | $ & $ n_{p2} = |S_p \cap T_2 | $ &  $ \hdots $ & $ n_{pq} = | S_p \cap T_q | $ &  $ n_{p\cdot} = | S_p | $\\ \cmidrule{2-7}
        & &  $ n_{\cdot 1} = |T_1| $ &  $ n_{\cdot 2} = |T_2| $ &  $ \hdots $ & $ n_{\cdot q} = |T_q| $ &  $ n $\\
    \end{tabular}
\end{table}

A partition loss function is computed from a contingency table as shown in Table
\ref{tab:contingency}, i.e., a cross-tabulation of counts among all pairs of
subsets from two partitions $\rho = \{S_1,\ldots,S_p\}$ and $\hat\rho =
\{T_1,\ldots,T_q\}$. Conceptually, we think of $\rho$ as the true partition and
$\hat\rho$ as being its estimate.  At times it is convenient to use the
equivalent representation with cluster labels, where we use $\bm{c} =
(c_1,\ldots,c_n)$ to denote the population clustering and $\hat{\bm{c}} =
(\hat{c}_1,\ldots,\hat{c}_n)$ to be its estimate.

\subsection{Binder Loss and Related Criteria}\label{subsec:binder}

\citet{binder.1978} loss has historically been the most widely used
loss function to estimate a random partition. He suggested the following:
\begin{equation}\label{binder}
L_{\text{binder}}(\textbf{c}, \hat{\textbf{c}}) = \sum_{i<j} \left( a\cdot\mathbb{I}\{c_i = c_j\}\mathbb{I}\{\hat{c}_i \neq \hat{c}_j\} + b\cdot\mathbb{I}\{c_i \neq c_j\}\mathbb{I}\{\hat{c}_i = \hat{c}_j\}\right),
\end{equation}
where $a > 0$ and $b > 0$ give the unit costs for pairwise misclassification.
Specifically, $a$ represents the cost of failing to cluster together two items
which should in fact be clustered together, whereas $b$ represents the cost of
clustering two items which should in fact be separate.  \citet{lau.green.2007}
noted that minimizing the posterior expectation of Binder loss is equivalent to
maximizing $ f(\hat{\textbf{c}}) = \sum_{i<j} \mathbb{I} \{ \hat{c}_i =
\hat{c}_j\} (\pi_{ij}-b/(a+b) ),$ where the posterior similarity matrix
$\bm{\pi}$ is an $n$-by-$n$ matrix with elements $\pi_{ij} = \text{Pr}(c_i = c_j
\mid \mathcal{D})$.  Its Monte Carlo estimate $\hat{\bm{\pi}}$ is obtained using
$\hat\pi_{ij} = \frac{1}{H}\sum_{h=1}^H \mathbb{I} \{ c_i^{(h)} = c_j^{(h)} \}$,
which can be computed before optimization.  The task then becomes to maximize $
f(\hat{\textbf{c}}) = \sum_{i<j} \mathbb{I} \{ \hat{c}_i =
\hat{c}_j\}\Big(\hat{\pi}_{ij}-b/(a+b) \Big).$

Without a reference to loss functions, \citet{dahl.2006} suggested the ``least
squares clustering'' criterion which seeks the clustering that minimizes $
f(\hat{\textbf{c}}) = \sum_{i=1}^{n} \sum_{j=1}^{n}(A(\hat{\textbf{c}})_{ij} -
\hat{\pi}_{ij})^2,$ where $A(\hat{\textbf{c}})$ is an $n$-by-$n$ adjacency
matrix whose $ij^\text{th}$ element is $\mathbb{I}\{ \hat{c}_i = \hat{c}_j \}$.
\citet{dahl.newton.2007} noted that minimizing this criterion is equivalent to
minimizing the Monte Carlo estimate of the posterior expectation of Binder loss when
$a=b$.

\citet{wade.ghah.2018} introduced an ``$n$-invariant version'' of Binder loss,
which is interpretable across sample size $n$, under the assumption that
$a=b=1$:
\begin{equation}\label{binder_n_invariant}
L_{\text{n'binder}}(\rho, \hat\rho) = \frac{2}{n^2} L_{\text{binder}}(\rho, \hat\rho)= \sum_{S \in \rho} \left( \frac{|S|}{n}\right)^2 + \sum_{T \in \hat{\rho}} \left( \frac{|T|}{n}\right)^2 - 2\sum_{S \in \rho} \sum_{T \in \hat{\rho}} \left( \frac{|S \cap T |}{n} \right)^2.
\end{equation}
Although Binder loss was first introduced with general costs $a$ and $b$ of
pairwise misclassification, to our knowledge, every software implementation
forces $a=b$ and applications seem to invariably assume that $a=b$.  Whereas the
$n$-invariant Binder loss of \citet{wade.ghah.2018} assumes $a=b=1$, we provide
in Section \ref{sec:generalized} an $n$-invariant version of Binder loss that
restores the flexibility of potentially different costs $a$ and $b$ in (\ref{binder}) and is
computationally no harder to evaluate. We demonstrate the advantage of this
flexibility in Section \ref{sec:verifications}, noting that the often criticized
property of Binder loss overestimating the number of clusters is merely a result
of assuming $a=b$, which is easily remedied by using $a > b$.

\citet{rand.1971} introduced a measure of similarity between two partitions,
which can be expressed in terms of Binder loss with $a=b=1$ as $ RI(\rho,
\hat\rho) = 1-L_{\text{binder}}(\rho, \hat\rho) / \binom{n}{2}.$ Maximizing the
posterior expectation of the Rand index (RI) is equivalent to minimizing the
posterior expectation of Binder loss with $a=b=1$. \citet{hubert.arabie.1985}
proposed the adjusted Rand index which accounts for chance agreement, yielding:
\begin{equation}\label{ari_matrix}
AR(\rho, \hat\rho) = \frac{\sum_{S \in \rho} \sum_{T \in \hat\rho} {|S \cap T| \choose 2} - \sum_{S \in \rho}{|S|\choose 2}\sum_{T \in \hat\rho} {|T|\choose 2}/{n\choose 2}}
{\frac{1}{2}\left[ \sum_{S \in \rho} {\binom{|S|}{2}} + \sum_{T \in \hat\rho} {\binom{|T|}{2}}\right] - \sum_{S \in \rho}{\binom{|S|}{2}}\sum_{T \in \hat\rho} {\binom{|T|}{2}}/{\binom{n}{2}}}.
\end{equation}
As large values of the adjusted Rand index indicate more similarity between
clusterings, \citet{fritsch.ickstadt.2009} entertained the idea of seeking the
clustering that maximizes the posterior expectation of the adjusted Rand index.
However, they found it computationally expedient to instead maximize an
approximation of that expectation:
\begin{equation}\label{pear_2}
f(\hat{\textbf{c}}) = \frac{\sum_{i<j} \mathbb{I}\{\hat{c}_i = \hat{c}_j\} \hat{\pi}_{ij} - \sum_{i<j} \mathbb{I}\{\hat{c}_i = \hat{c}_j\} \sum_{i<j} \hat{\pi}_{ij}/{n\choose 2}}
{\frac{1}{2}\left[\sum_{i<j} \mathbb{I}\{\hat{c}_i = \hat{c}_j\} + \sum_{i<j} \hat{\pi}_{ij}\right] - \sum_{i<j} \mathbb{I}\{\hat{c}_i = \hat{c}_j\} \sum_{i<j} \hat{\pi}_{ij}/{n\choose 2}}.
\end{equation}
Using our computational techniques that we detail in Section \ref{sec:salso}, we
find that resorting to the approximation is not necessary and, for large $n$,
may be detrimental to computations.  Instead, we suggest maximizing the
posterior expectation of the adjusted Rand index, or equivalently minimizing the
posterior expectation of ``one minus the adjusted Rand index'' (omARI) loss,
i.e., $ L_{\text{omARI}}(\rho, \hat\rho) = 1 - AR(\rho, \hat\rho). $

\subsection{Variation of Information and Other Information-Based Losses}\label{subsec:vi}

An alternative, more recently introduced class of partition loss functions based
on information theory has been proposed and studied by \citet{meila.2007},
\citet{vinh.etal.2010}, \citet{wade.ghah.2018}, and \citet{rastelli.friel.2018}.
 These are calculated using the totals from the contingency table (see Table
\ref{tab:contingency}) and are functions of individual entropies $H(\rho)$ and
$H(\hat\rho)$, joint entropy $H(\rho, \hat\rho)$, and mutual information
$I(\rho, \hat\rho)$, as defined below:
\begin{equation}
\label{entropy_c}
\begin{split}
H(\rho) = -\sum_{S \in \rho} \frac{|S|}{n} \log_2\left( \frac{|S|}{n}\right) \quad & \quad \quad  H(\rho, \hat\rho) = -\sum_{S \in \rho} \sum_{T \in \hat{\rho}} \frac{|S \cap T |}{n} \log_2\left( \frac{|S \cap T |}{n} \right) \\
H(\hat\rho) = -\sum_{T \in \hat{\rho}} \frac{|T|}{n} \log_2\left( \frac{|T|}{n}\right) \quad & \quad \quad I(\rho, \hat\rho) = H(\rho) + H(\hat\rho) - H(\rho, \hat\rho)
\end{split}
\end{equation}
Note the use of the binary logarithm $\log_2(\cdot)$.  The conditional entropy
$H(\rho\mid \hat{\rho})$ can be written in several forms, including $H(\rho\mid
\hat{\rho}) = H(\rho,\hat{\rho}) - H(\hat{\rho}) = H(\rho) -
I(\rho,\hat{\rho})$.

The variation of information (VI) was introduced by \citet{meila.2007} as a
measure of distance between two partitions. \citet{wade.ghah.2018} were the
first to consider using the VI as a loss function to estimate a random partition.  The
VI is expressed as:
\begin{align}\label{vi}
L_{\text{VI}}(\rho, \hat\rho) &= H(\rho) + H(\hat\rho) - 2I(\rho, \hat\rho)\\
&= -H(\rho) - H(\hat\rho) + 2H(\rho, \hat\rho) \nonumber\\
&= \sum_{S \in \rho} \frac{|S|}{n} \log_2\left( \frac{|S|}{n}\right) + \sum_{T \in \hat{\rho}} \frac{|T|}{n} \log_2\left( \frac{|T|}{n}\right) - 2\sum_{S \in \rho} \sum_{T \in \hat{\rho}} \frac{|S \cap T |}{n} \log_2\left( \frac{|S \cap T |}{n} \right).\nonumber
\end{align}
Since $H(\rho)$ is constant when minimizing with respect to $\hat\rho$,
\citet{wade.ghah.2018} note that minimizing the posterior expectation of VI loss
is equivalent to minimizing:
\begin{align}\label{optimize_vi}
\hat{\textbf{c}}^* &= \underset{\hat{\textbf{c}}}{\text{argmin}}\hspace{3pt}\mathbb{E}(L_{\text{VI}}(\textbf{c}, \hat{\textbf{c}}) \mid \mathcal{D})\\
&= \underset{\hat{\textbf{c}}}{\text{argmin}}\sum_{i=1}^{n}\log_2\left( \sum_{j=1}^{n} \mathbb{I}(\hat{c}_{j} = \hat{c}_j)\right) - 2\sum_{i=1}^{n}\mathbb{E}\left( \log_2\left( \sum_{j=1}^{n}\mathbb{I}(c_{j}=c_i)\mathbb{I}(\hat{c}_{j} = \hat{c}_i)\right) \mid \mathcal{D}\right)\nonumber.
\end{align}
The expectation at the end of (\ref{optimize_vi}) can be approximated using
Monte Carlo integration based on posterior samples
$\bm{c}^{(1)},\ldots,\bm{c}^{(H)}$ obtained from MCMC.  \citet{wade.ghah.2018}
note, however, that evaluating this Monte Carlo approximation is $O(Hn^2)$,
which may be costly when considering many candidate $\hat{\bm{c}}$'s in an
optimization procedure.  As such, they suggest applying Jensen's inequality to
swap the logarithm and expectation, yielding a lower bound on
(\ref{optimize_vi}). Specifically, \citet{wade.ghah.2018} suggest:
\begin{equation}\label{optimize_vilb}
\hat{\textbf{c}}^* = \underset{\hat{\textbf{c}}}{\text{argmin}}\sum_{i=1}^{n}\log_2\left( \sum_{j=1}^{n} \mathbb{I}(\hat{c}_{j} = \hat{c}_i)\right) - 2\sum_{i=1}^{n}\left( \log_2\left( \sum_{j=1}^{n}\hat{\pi}_{ij}\mathbb{I}(\hat{c}_{j} = \hat{c}_i)\right)\right),
\end{equation}
where $\hat\pi_{ij} = \frac{1}{H}\sum_{h=1}^H \mathbb{I} \{ c_i^{(h)} =
c_j^{(h)} \}$ can be cached before optimization, making the computational
complexity $O(n^2)$ for a given $\hat{\bm{c}}$. In seeking to minimize the
\emph{lower bound} of the posterior expectation of VI, \citet{wade.ghah.2018}
approach VI in an analogous manner to how \citet{fritsch.ickstadt.2009} approach
omARI, in that both seek to optimize an \emph{approximation} to the posterior
expectation of the target quantity. The effect of applying Jensen's inequality
in this case has not been formally investigated, although our experience is that
minimizing the two criteria almost never leads to the same estimate in typical
applications. More to the point, although \citet{meila.2007} and
\citet{vinh.etal.2010} find many desirable properties of VI, the extent to which
they still hold when applying Jensen's inequality is not well understood.

In Section \ref{sec:generalized} we provide a generalization of the original
variation of information with weights $a$ and $b$ to influence the trade-offs
between (i) failing to cluster two items which should be together and
(ii) clustering two items which should be separate.  These weights are analogous
to those in Binder loss.  We show that our generalization maintains a desirable
property of the original VI, specifically that our generalization is still a
quasimetric. Moreover, whereas \citet{wade.ghah.2018} resort to an approximation
of the posterior expectation of VI, we show in Section \ref{sec:salso} a
computationally-cheap way to evaluate the actual expectation, whether for the
original VI or for our generalization with unequal $a$ and $b$.

\citet{vinh.etal.2010} consider two dozen variants of the information-based
distances and their respective properties. As they explain, the normalized
variation of information (NVI) takes the VI as defined in (\ref{vi}), which has
a range of $[0,\log_2{n}]$, and scales it to have a range of $[0,1]$. The NVI is
defined as $ L_{\text{NVI}}(\rho, \hat\rho) = 1-\frac{I(\rho, \hat\rho)}{H(\rho,
\hat\rho)}.$ Instead of the NVI, however, \citet{vinh.etal.2010} advocate for
the normalized information distance (NID) as a general purpose loss function
with useful and important properties, defined as $ L_{\text{NID}}(\rho,
\hat\rho) = 1-\frac{I(\rho, \hat\rho)}{\max\{H(\rho), H(\hat\rho)\}}.$ The NID
is normalized in that it takes values in $[0,1]$; there is a corresponding
unnormalized version taking values in $[0,\log_2{n}]$, which we call the
information distance (ID), defined as $ L_{\text{ID}}(\rho, \hat\rho) =
\max\{H(\rho), H(\hat\rho)\} - I(\rho, \hat\rho).$

All of the loss functions described in this section can be used with our SALSO
algorithm that we present in Section \ref{sec:salso}. We compare these loss
functions in Section \ref{sec:verifications}.

\section{Generalizations of Binder and VI Loss}
\label{sec:generalized}

\subsection{Generalization of Binder Loss}

A major criticism of the Binder loss function in (\ref{binder}) is that it tends
to overestimate the number of clusters
\citep{wade.ghah.2018,rastelli.friel.2018}, although these results were obtained
with the implicit assumption that $a=b$. Recall that the Binder loss function
has weights $a$ and $b$ which are positive constants, where $a$ represents the
cost of failing to cluster two items which should be together and $b$
represents the cost of clustering two items which should be separate.  With the
notable exception being \citet{lau.green.2007}, the original statement of Binder
loss with potentially unequal weights has been lost in the literature. This is
probably due to the fact that most publicly available software implementations
use Binder loss with equal weights, except for our \texttt{salso} package.

Thus, the criticism of the Binder loss function overestimating the number of
clusters may well be an artifact that practitioners are using equal costs of
misclassification. Intuitively, one would expect that using $a > b$ would
address the overestimation of the number of clusters, because this defines a
greater cost to incorrectly split clusters up than there is to incorrectly
cluster more items together.  Taken to the extremes, for fixed $b$, it is
evident that $a\rightarrow0$ yields a clustering estimate with every item in a
singleton cluster and that $a \rightarrow \infty$ yields a clustering estimate
with every item in the same cluster.

Recall that \citet{wade.ghah.2018} provide (\ref{binder_n_invariant}), an
expression for the Binder loss in terms of the contingency table when $a=b$.
Expressing the Binder loss in terms of the contingency table facilitates fast
computations in exploring the partition space, as we describe in Section
\ref{sec:salso}.  Extending the work of \citet{wade.ghah.2018}, below we provide
an expression for the $n$-invariant Binder loss in terms of general weights $a$
and $b$:
\begin{dfn}
	For weights $a,b > 0$, the $n$-invariant Binder loss is:
	\begin{equation}
	\label{generalizedBinderNinvariant}
	L_{\text{n'binder}}(\rho,\hat\rho) = a \sum_{S \in \rho} \left( \frac{|S|}{n}\right)^2 + b \sum_{T \in \hat{\rho}} \left( \frac{|T|}{n}\right)^2 - (a+b) \sum_{S \in \rho} \sum_{T \in \hat{\rho}} \left( \frac{|S \cap T |}{n} \right)^2.
	\end{equation}
\end{dfn}
We refer to (\ref{generalizedBinderNinvariant}) as the generalized Binder loss.
We do this to contrast it with (\ref{binder_n_invariant}) and to differentiate
it from the typical practice in the literature. A proof of the equivalence
between the Binder loss as it was originally stated in (\ref{binder}) and
(\ref{generalizedBinderNinvariant}) in terms of the counts is found in the
Appendix. We show in Section \ref{sec:verifications} that this generalized
Binder loss with $a > b$ has the desired property of controlling the number of
clusters in the estimate.

We believe that some of the rise in popularity of the variation of information
loss, which tends to produce fewer clusters, is due to the use of Binder loss with
equal weights. We hope that noting the utility of unequal weights together with
the fast software implementation in \texttt{salso} has the potential to reignite
interest in the Binder loss function.

\subsection{Generalization of the Variation of Information}

While the variation of information tends to yield few clusters, it has been
suggested that in some cases the variation of information may produce
\textit{fewer} clusters than is reasonable.
In order to counteract this, we propose the generalized variation of information
loss (GVI), with positive weights $ a $ and $ b $ which has the following form:
\begin{dfn}
	For weights $a,b > 0$, the generalized variation of information (GVI) is:
	\begin{align}\label{gvi_def}
L&_{\text{GVI}}(\rho, \hat\rho) = bH(\rho) + aH(\hat\rho) - (a+b)I(\rho, \hat\rho)\\
&= -aH(\rho) - bH(\hat\rho) + (a+b)H(\rho, \hat\rho)\nonumber\\
&= a\sum_{S \in \rho} \frac{|S|}{n} \log_2\left( \frac{|S|}{n}\right) + b\sum_{T \in \hat{\rho}} \frac{|T|}{n} \log_2\left( \frac{|T|}{n}\right) - (a+b)\sum_{S \in \rho} \sum_{T \in \hat{\rho}} \frac{|S \cap T |}{n} \log_2\left( \frac{|S \cap T |}{n} \right).\nonumber
\end{align}
\end{dfn}
Note the similarity between this proposed measure and Binder loss as defined in
(\ref{generalizedBinderNinvariant}). We hypothesized that this generalized
variation of information would allow control over the number of clusters in
estimates similar to the control allowed by the generalized Binder loss.  We
show in Section \ref{sec:verifications} that this is in fact the case with $a<b$
(e.g., $a=0.5, b=1$). While it is not possible to write (\ref{gvi_def}) in terms
of the sum of per-unit costs --- as can be done for the Binder loss in
(\ref{binder}) --- it is nonetheless the case that GVI behaves similarly to the
Binder loss in that, for fixed $b$, $a\rightarrow 0$ yields a clustering
estimate with every item in a singleton cluster and $a \rightarrow \infty$
yields a clustering estimate with every item in the same cluster.

A desirable property of a loss function is whether it is a metric
\citep{vinh.etal.2010}. In order to satisfy the metric property, a distance
measure must satisfy the following three properties: 1) the identity of
indiscernibles, 2) symmetry, and 3) the triangle inequality. Quasimetrics are
distance measures with the identity of indiscernibles and the triangle
inequality, but not symmetry. \citet{meila.2007} and \citet{wade.ghah.2018}
proved the metric property for Binder loss assuming $a=b$, but the properties of
the generalized Binder loss ($ a\neq b $) are not discussed in the literature.
We thus provide the following theorem:
\begin{thm}\label{gbinder_quasimetric}
	The generalized Binder loss is a quasimetric.
\end{thm}
Likewise, \citet{meila.2007} and \citet{vinh.etal.2010} proved the metric
property for the variation of information. Similarly, the following holds for
GVI:
\begin{thm}\label{gvi_metric}
	The generalized variation of information (GVI) is a quasimetric.
\end{thm}
We provide proofs for both theorems in the Appendix.

\section{Existing Algorithms for Partition Summarization}
\label{sec:algorithms}

Computing a Monte Carlo estimate of the posterior expected loss requires both a
loss function --- whether an existing one from Section \ref{sec:criteria} or one
of our extensions from Section \ref{sec:generalized} --- and samples from the
posterior partition distribution.  A far more challenging practical problem,
however, is searching the vast space of partitions for the minimizer of the
chosen criterion, i.e., searching for the minimizer of the Monte Carlo estimate
of the expectation of the chosen loss function.  We review the literature of
existing search algorithms in this section and then, in Section \ref{sec:salso},
propose SALSO, a new search algorithm for any loss function that is
substantially faster and leads to demonstrably better estimates than existing
algorithms, yet has tractable complexity such that it readily scales in $n$
items.

Obviously an exhaustive search of all possible clusterings of $n$ items,
evaluating the chosen criterion for each clustering, will yield the absolute
minimizer.  Exhaustive enumeration is only feasible for very small $n$ because
the $n^{\text{th}}$ Bell number $B(n)$, i.e., the number of possible clusterings
of $n$ items, grows exponentially.  For example, $B(50)$ is more than $10^{47}$.
\citet{dahl.2006} suggested the draws method which simply selects, among all
those in the MCMC output, the clustering that minimizes the chosen criterion.
This method is practical, fast, and applicable to any loss function, but the
clustering estimate is obviously limited to the clusterings visited by the
Markov chain.  We will now review existing methods that can produce clustering
estimates beyond those visited by the Markov chain.

\subsection{\citet{medv.siva.2002} - Hierarchical Clustering}

Without reference to any loss function, \citet{medv.siva.2002} proposed using
hierarchical agglomerative clustering based on MCMC output. Hierarchical
clustering uses a distance matrix, for which \citet{medv.siva.2002} proposed one
minus the estimated posterior similarity matrix $\hat{\bm{\pi}}$ defined in
Section \ref{subsec:binder}. \citet{medv.siva.2002} used complete linkage, but
other linkage methods could be used. This method has been viewed as \textit{ad
hoc} \citep[see][]{dahl.2006, fritsch.ickstadt.2009} because it is not based on
a loss function and it builds a full tree of possible clusterings, leaving the
problem of where to cut the tree. This method does, however, quickly give a
reasonable answer.

\citet{fritsch.ickstadt.2009} took this idea further in the \texttt{R} package
\texttt{mcclust} \citep{pkg.mcclust}, where they cut the tree (i.e., select the
clustering among those implied by the tree) to minimize the Monte Carlo estimate
of the posterior expected loss, yielding a clearly defined implementation based
on a loss function.  Nevertheless, \cite{rastelli.friel.2018} found in a
simulation study comparing several procedures that ``it is clear that the
[\citet{medv.siva.2002} method with cuts from \citet{fritsch.ickstadt.2009}]
performs quite poorly'' in terms of the quality of the clustering estimate
produced.

\subsection{\citet{lau.green.2007} - Binary Integer Programming}

\citet{lau.green.2007} proposed two search procedures for an optimal clustering.
The first involves formulating and solving a binary integer programming problem,
but is impractical as $n$ increases. Recognizing the intractability,
\citet{lau.green.2007} propose another procedure, a heuristic item-swapping
algorithm as a fast approximation of their first procedure. The second
procedure, however, still suffers from scalability problems.
\citet{rastelli.friel.2018} state that ``the method of \citet{lau.green.2007}
scales very poorly with $n$,'' as seen in Figure 1 of their paper. Likewise,
\citet{fritsch.ickstadt.2009} were able to apply the algorithm for $n=200$ after
several hours of computation, but found ``it was not possible to apply the
algorithm to all 400 observations, as the optimization problems required at each
iteration got too large to be handled by the software.''  In addition to this
lack of scalability and slow computation, the methods proposed by
\citet{lau.green.2007} are only detailed in terms of Binder loss and the
estimated posterior similarity matrix.  It is not clear how broadly the
algorithm could be extended since, for example, the variation of information
cannot be expressed as a function of an estimated posterior similarity matrix.

\subsection{\citet{wade.ghah.2018} - Greedy Algorithm}

The method of \citet{wade.ghah.2018} is one of three ``greedy algorithms''
described in this paper. A greedy algorithm is a procedure that takes small,
locally-optimal updates at each step of the algorithm along the way to finding
its final solution. The method of \citet{rastelli.friel.2018} and our proposed
SALSO algorithm are also greedy algorithms.

The greedy search algorithm of \citet{wade.ghah.2018} takes locally-optimal
moves in a neighborhood of partitions defined in terms of the chosen loss
function and the Hasse diagram, a lattice in which the nodes are all possible
partitions and edges are those partitions that are one change away from each
other.  See \citet{wade.ghah.2018} for details and examples of the Hasse
diagram. Their algorithm is implemented for Binder loss and the \emph{lower
bound} of VI loss in the \texttt{mcclust.ext} package for \texttt{R}, available
on Wade's website. It is worth noting that their implementation relies on the
estimated posterior similarity matrix, and thus cannot be easily applied to loss
functions, such as VI, NID, and NVI.

One downside to this method is its dependence on the initial partition,
especially since the algorithm can get stuck in a local minimum as it traverses
the Hasse diagram. Even though it scales better in $n$ than the method proposed
by \citet{lau.green.2007}, \citet{rastelli.friel.2018} conclude from their
simulation study that ``the method of Wade and Ghahramani does not scale
particularly well.''

\subsection{\citet{rastelli.friel.2018} - Greedy Algorithm}
\label{sec:rastelli.frield.2018}

The greedy search algorithm of \citet{rastelli.friel.2018} starts at a
randomly-selected partition with many small clusters and iteratively reassigns
one item at a time to existing clusters or a new singleton cluster, where
reassignment decisions are made to minimize the Monte Carlo estimate of the
expected loss.  A scan is completed once each item, in a random order, has been
considered for reassignment and the algorithm stops once a scan yields no
change.

The search algorithm proposed by \citet{rastelli.friel.2018} is stochastic in
nature, as the starting partition is assigned randomly and the one-at-a-time
optimizations are done in a random order each time. Although
\citet{rastelli.friel.2018} recommend against multiple runs of their algorithm,
we show in Section \ref{sec:verifications} that multiple runs can
greatly increase the chance of a better answer and that, for difficult problems,
it is very unlikely that the optimal partition will be obtained on only one run.

The algorithm is implemented by the authors in the \texttt{GreedyEPL} package (which is archived on
CRAN) for the following losses: Binder, VI, NVI, and NID.  We note that the user
can specify the maximum number of clusters $K_{\text{up}}$, but the software
will actually use the maximum of $K_{\text{up}}$ and the largest number of
clusters found in any iteration of the MCMC output. Thus, while the method as
described in the paper can control the maximum number of clusters, in practice
their software can be unwieldy in this regard.

\section{SALSO Algorithm}
\label{sec:salso}

In this section, we describe our SALSO algorithm, investigate its complexity,
and discuss computational shortcuts such that
(\ref{eq_posterior_expectation_via_MC}) need not be fully evaluated for each
partition that is considered in the SALSO algorithm.

\subsection{Description of the SALSO Algorithm}

The SALSO algorithm, like most algorithms discussed in Section
\ref{sec:algorithms}, tries to perform the optimization in
(\ref{eq_generic_loss}) using Monte Carlo estimates in
(\ref{eq_posterior_expectation_via_MC}).  Like the method of
\citet{rastelli.friel.2018}, the SALSO algorithm is a greedy, stochastic search
and, with particular choices for its parameters, can mimic the behavior of the
Rastelli and Friel (R\&F) algorithm.  There are four phases of the SALSO
algorithm: initialization, sweetening, zealous updates, and recording. We
describe each below.  The algorithm is embarrassingly parallel and we advocate
for performing multiple runs. The algorithm is implemented for several loss
functions in the \texttt{salso} package available on CRAN.

The SALSO algorithm provides two methods to initialize a partition. The first
method is sequential allocation, in which each item is allocated one at a time
--- in the order determined by a permutation sampled uniformly among all
possible permutations --- to an existing cluster or a new cluster, based on the
allocation which minimizes the Monte Carlo estimate of the posterior expected
loss, ignoring any yet-to-be allocated items.  For example, the first
randomly-selected item is placed in a cluster by itself. The second
randomly-selected item is placed in either the cluster with the first item, or
in a new cluster by itself, depending on which minimizes the posterior expected
loss, computed as if there were only those two items in the system. The process
continues and the number of clusters may grow, although we never consider adding
a new cluster if that would lead to a partition having more than the desired
maximum number of clusters, denoted $k_d$.  In the second initialization method,
cluster labels are obtained by uniformly sampling the labels $1, \ldots, k_d$.
The initialization method for a particular run is randomly chosen and the user
can specify the probability of sequential allocation, denoted $p_{\text{SA}}$,
with $0.5$ being its default value.

\begin{algorithm}
\small
  \caption{\small Pseudocode for the SALSO Algorithm\label{salsoAlgorithm}.  Let
  $n_\text{runs}$ be the number of runs, $p_\text{SA}$ be the probability of
  sequential allocation, $k_d$ be the maximum number of clusters, and
  $n_\text{maxZealous}$ be the maximum number of zealous updates. Note that,
  when considering ``all existing clusters and a new cluster'' below, the
  algorithm does \emph{not} consider a new cluster if that would lead to a state
  in which the number of clusters exceeds the maximum number of clusters $k_d$.}
  \label{algorithm:salso}
  \begin{algorithmic}[1]
    \For{$i = 1,...,n_{\text{runs}}$}
        \State \Comment{Initialization phase}
        \If{Uniform[0,1] $<$ $p_{\text{SA}}$}\Comment{Do sequential initialization}
            \State Uniformly sample a permutation $(\sigma_1,...,\sigma_n)$ of $ \{1,...,n\}$.
            \For{$ \sigma = \sigma_1, \ldots, \sigma_n $}
                \For{$c$ being all existing clusters and a new cluster}
                    \State Calculate the posterior expected loss if $\sigma$ were added to cluster $c$.
                \EndFor
                \State Allocate $\sigma$ to the cluster that minimizes the posterior expected loss.
            \EndFor
        \Else \Comment{Do random initialization}
            \State Sample a random partition with at most $k_d$ clusters,
            \State \hspace{3ex} e.g., \texttt{sample(1:$k_d$, $n$, replace=TRUE)} in \texttt{R}.
        \EndIf
        \State \Comment{Sweetening phase}
        \While{\texttt{TRUE}}
            \State Uniformly sample a permutation $(\sigma_1,...,\sigma_n)$ of $ \{1,...,n\}$.
            \For{$\sigma = \sigma_1, \ldots, \sigma_n$}
                \State Remove $\sigma$ from its cluster.
                \For{$c$ being all existing clusters and a new cluster}
                    \State Calculate the posterior expected loss if $\sigma$ were added to cluster $c$.
                \EndFor
                \State Allocate $\sigma$ to the cluster that minimizes the posterior expected loss.
        \EndFor
        \If{partition is unchanged} \textbf{break}
        \EndIf
        \EndWhile
        \State \Comment{Zealous updates phase}
            \For{$c^\prime$ in a random ordering of up to $n_{\text{maxZealous}}$ clusters}
                \State Record the current partition as $\hat{\rho}$.
                \State Destroy cluster $c^\prime$ by removing (deallocating) all of its items.
                \For{$\sigma$ in a random ordering of deallocated items}
                    \For{$c$ being all existing clusters and a new cluster}
                        \State Calculate the posterior expected loss if $\sigma$ were added to cluster $c$.
                    \EndFor
                    \State Allocate $\sigma$ to the cluster that minimizes the posterior expected loss.
                \EndFor
                \If{$\hat{\rho}$ has a smaller posterior expected loss than the current partition}
                    \State Revert the current partition to $\hat{\rho}$.
                \EndIf
            \EndFor
        \State \Comment{Recording phase}
        \State Record the current partition as $\hat{\rho}_i$ and note its posterior expected loss.
    \EndFor
    \State \Return the $\hat{\rho}_i$ with the smallest posterior expected loss.
  \end{algorithmic}
\end{algorithm}

Once the partition is initialized, the next step is the sweetening phase, in
which random one-at-a-time reallocations of individual items are performed in a
random order. This is essentially the same idea as in the sequential allocation
method, except now every item is allocated and each item --- one at a time and
in the order determined by a permutation sampled uniformly among all possible
permutations --- is removed from its cluster and reallocated to existing clusters
or a new cluster, according to the choice that minimizes the Monte Carlo
estimate of the posterior expected loss.  This process is repeated until there
is no change after a complete pass through all $n$ items.

In the third phase, ``zealous'' updates attempt to break out of a local minimum.
 As these zealous updates can be computationally expensive, we suggest setting
an upper bound on the number of such updates, denoted $n_\text{maxZealous}$.
(The default in our software is 10.)  For a random ordering of up to
$n_\text{maxZealous}$ clusters, the current state is recorded and the cluster is
destroyed by removing all of its items. These deallocated items are then
sequentially reallocated --- one at a time and in a random order ---
conditioning on the already allocated items, in the same way as sequential
allocation in the initialization phase.  Once everything is reallocated, the
Monte Carlo estimate of the posterior expected loss of the current partition is
compared to what it was before destroying the cluster. If no improvement was
found by this zealous update, it is abandoned and the state reverts to the
previous state.

Finally, in the fourth phase, the Monte Carlo estimate of the posterior expected
loss is recorded for the current state.  This algorithm is ``embarrassingly
parallel'' --- since each run of the algorithm does not rely on any other run
--- such that $n_\text{runs}$ runs of the algorithm can easily be conducted
using all available CPU cores.  Among all $n_\text{runs}$ candidates, the
partition with the smallest Monte Carlo estimate is then reported as the
partition estimate. The entire SALSO algorithm is shown in pseudocode in
Algorithm \ref{salsoAlgorithm}.

We note that our SALSO algorithm almost reduces to that of
\citet{rastelli.friel.2018} when: i.\ the probability of sequential allocation
$p_\text{SA}$ is set to $0$, ii.\ the number of zealous updates
$n_\text{maxZealous}$ is set to 0, and iii.\ the number of runs $n_\text{runs}$
is set to 1.  Even with those specific choices, however, significant practical
differences remain between these two algorithms.  First, the default (and
recommended) value for the maximum number of clusters $k_d$ in
\citet{rastelli.friel.2018} is $n$, but this can worsen complexity, greatly slow
down computations, increase RAM requirements, and lead to uninterpretable
estimates. (See Section \ref{sec:complexity} and Section
\ref{sec:verifications}.) Further, their implementation forces $k_d$ to be no
smaller than the maximum number of clusters observed among the posterior samples
$\rho^{(1)},\ldots,\rho^{(H)}$, which still may result in too many clusters for
interpretable estimates.  In our implementation, the user has full control over
$k_d$ and we set its default value to be the maximum number of clusters observed
instead of $n$.  Another practical difference is that R\&F ``hard-coded''
epsilon in their stopping rule in the sweetening phase, which does not account
for differences in scales among various loss functions.  Finally, we show in
Section \ref{sec:verifications} that our implementation is faster, in part
because of the computational shortcuts that we detail in Section
\ref{sec:shortcuts}.

\subsection{Complexity Comparison}\label{sec:complexity}

We now compare the complexity of the SALSO algorithm with those of
\citet{wade.ghah.2018} and \citet{rastelli.friel.2018} in greater detail. The
complexity of the SALSO algorithm is $\mathcal{O}(H \cdot k_d \cdot k_H \cdot
n)$, where $H$ is the number of MCMC samples, $k_d$ is the maximum number of
clusters desired by the user, $k_H$ is the maximum number of clusters observed
among the MCMC samples, and $n$ is the number of items. We recommend setting
$k_d$ to a relatively small number for the sake of interpretion of the
clustering estimate. It defaults to $k_H$, which is typically much smaller
than $n$. So, the default complexity for SALSO is $\mathcal{O}(H \cdot k_H^2
\cdot n)$, but it will be less if the user specifies $k_d < k_H$. The complexity
of the R\&F algorithm is $\mathcal{O}(H \cdot k_d^2 \cdot n)$, and the
implementation defaults to $k_d$ being $n$ and requires that $k_d$ be at least
$k_H$. So, in the default case, the R\&F algorithm has complexity $\mathcal{O}(H
\cdot n^3)$ and, in the best case, has complexity $\mathcal{O}(H \cdot k_H^2
\cdot n)$. Finally, the complexity of the algorithm of \citet{wade.ghah.2018} is
$\mathcal{O}(\ell \cdot n^2) $, where $2\ell$ defines the number of partitions
to consider at each iteration. \citet{wade.ghah.2018} recommend a default value
of $\ell=n$, meaning that their default complexity becomes $\mathcal{O}(n^3)$.

\subsection{Computational Speedups}
\label{sec:shortcuts}

Notice that on lines 6-8, 17-19, and 26-28 in Algorithm \ref{algorithm:salso},
the SALSO algorithm needs to allocate the current item, denoted $\sigma$, to one
of the existing clusters or to a new cluster and that this choice is made such
that the Monte Carlo estimate of the posterior expected loss is minimized. Let
$\hat{\bm{c}}_1, \ldots, \hat{\bm{c}}_q, \hat{\bm{c}}_{q+1}$ denote the
clusterings obtained by allocating the current item $\sigma$ to the $q$ existing
clusters or to a new cluster. To make this allocation, then, it would seem that
the SALSO algorithm must compute:
\begin{equation}
    \label{eq_posterior_expectation_via_MC_salso}
    \mathbb{E}(L(\bm{c}, \hat{\bm{c}}_j \mid \mathcal{D}) \approx \frac{1}{H} \sum_{h=1}^H L(\bm{c}^{(h)}, \hat{\bm{c}}_j), \quad \text{ for } j=1,\ldots, q, q+1,
\end{equation}
where $\bm{c}^{(1)},\ldots,\bm{c}^{(H)}$ are $H$ samples from a posterior
distribution $p(\bm{c} \mid \mathcal{D})$ and then allocate the current item,
$\sigma$, to the cluster among all $q+1$ explored that minimizes the Monte Carlo
estimate of the posterior expected loss, i.e., to cluster $s$ where
\begin{equation}
    \label{eq_minimizer_among_explored}
    \frac{1}{H} \sum_{h=1}^H L(\bm{c}^{(h)}, \hat{\bm{c}}_s) \ \le \ \frac{1}{H} \sum_{h=1}^H L(\bm{c}^{(h)}, \hat{\bm{c}}_j), \hspace{3ex} \forall \hspace{3pt} j\in \{1,\ldots, q, q+1\}.
\end{equation}
We have recognized and implemented certain computational speedups that have
allowed for the efficient repeated calculation of the Monte Carlo approximation
of the posterior expected loss. This is possible when the loss is written in
terms of counts, e.g, as in (\ref{generalizedBinderNinvariant}) and
(\ref{gvi_def}).  Both i. allocating a new item to a cluster and ii. moving an
item from one cluster to another require updating just four counts in a
contingency matrix (see Table \ref{tab:contingency}).
Because only four counts are updated, software can easily cache these
contingency tables --- one for each $\bm{c}^{(1)},\ldots,\bm{c}^{(H)}$ --- and
make only the four required updates per MCMC sample after each change to the
estimated partition. This allows for efficient incremental calculation and
storage of the contingency matrices necessary for calculation of the loss
functions.

For some loss functions, further shortcuts are possible.  For the generalized
Binder loss in (\ref{generalizedBinderNinvariant}), picking the $\hat{\bm{c}}_j$
with the smallest value in (\ref{eq_posterior_expectation_via_MC_salso}) is
equivalent to choosing among:
\begin{equation}
    \label{eq_posterior_expectation_via_MC_salsotwo}
    b H n_{\cdot j} - (a+b) \sum_{h=1}^H n_{c^{(h)} j}, \quad \text{ for } j=1,\ldots,q,q+1,
\end{equation}
where $n_{\cdot j}$ and $n_{ij}$ are defined in the contigency table (Table \ref{tab:contingency}) and $c^{(h)}$
is the cluster label of the current item $\sigma$ in the $h^{\text{th}}$ posterior sample.  Computing
($\ref{eq_posterior_expectation_via_MC_salsotwo}$) is very fast given cached
counts in a contingency table. We implement a similar shortcut for our
generalized variation of information (GVI) in (\ref{gvi_def}). That is, we choose among:
\begin{equation}
    \label{eq_posterior_expectation_via_MC_salsotwo_gvi}
    b H f( n_{\cdot j} ) - (a+b) \sum_{h=1}^H f ( n_{c^{(h)}j} ), \quad \text{ for } j=1,\ldots,q,q+1,
\end{equation}
where $f(n) = n \log_2(n) - (n-1)\log_2(n-1)$ is a function of an integer than
can be cached rather than repeatedly computed.

\section{Verifications}
\label{sec:verifications}

We now describe a simulation study to examine our SALSO algorithm using three
sets of samples from posterior partition distributions.  Each of these three
sets provide multiple model fits for different data or model specifications. In
our study, each procedure was replicated 10 times for each set of MCMC output
for a particular model.  We report results as averages across the 10
replications to mitigate dependence of the comparison on specific model choices,
oddities in any particular posterior sample, or random chance. These sets
cluster 60, 200, and 1,072 items respectively, allowing us to examine the
scalability of the algorithms in the number of items $n$. Our first set is
labeled ``PM10''.  \citet{page.etal.2020} analyze averaged monthly PM 10 data
from 60 stations in the European air quality database using 8 different models
for each of 12 months of data.  The PM10 set of posterior samples consists of a
collection of $8 \times 12 = 96$ models, each providing 1,000 samples for the
clustering of 60 stations. Our second set is labeled ``Gaskins'' and comes from
\citet{gaskins.etal.2017}, who perform a simulation study of many methods to
summarize samples from posterior clustering distributions using two sets of
posterior samples.  Here we consider the more challenging set, which consists of
200 distinct model fits, each providing 2,000 posterior samples, for the
clustering of 200 items. Our third set is labeled ``SIMCE'', which again comes
from \citet{page.etal.2020}.  They consider 8 models for 1,072 schools at each
of 7 time periods, yielding a collection of $8 \times 7 = 56$ models clustering
1,072 items with each model having 1,000 posterior samples.

\begin{table}[t]
    \footnotesize
    \centering
    \caption{\footnotesize The quality and run time of: (A) SALSO algorithm and
    (B) the default \citet{rastelli.friel.2018} algorithm from their software.
    Here both methods are allowed to produce results with any number of
    clusters.
        \newline
        \newline \textbf{A}: SALSO (10, 0.5), 1 run
        \newline \textbf{B}: R\&F 2018, 1 run\label{tab:simstudy1}}
    \begin{tabular}{llrrrrrr}
        \toprule
        & & \multicolumn{3}{c}{Quality} & \multicolumn{3}{c}{Run Time} \\
        \cmidrule(lr){3-5}
        \cmidrule(lr){6-8}
        Dataset & Loss & A $<$ B & B $<$ A & Diff. & A & B & Ratio \\
        \midrule
        PM10 & Binder & \textbf{0.14} & 0.01 & 0.13 & \textbf{0.01} & 0.03 & 0.41 \\
        & VI & \textbf{0.29} & 0.01 & 0.28 & \textbf{0.01} & 0.10 & 0.10 \\
        & NVI & \textbf{0.34} & 0.10 & 0.24 & \textbf{0.05} & 0.40 & 0.14 \\
        & NID & \textbf{0.32} & 0.10 & 0.23 & \textbf{0.05} & 0.30 & 0.15 \\
        \midrule
        Gaskins & Binder & \textbf{0.10} & 0.02 & 0.08 & \textbf{0.26} & 0.65 & 0.40 \\
        & VI & \textbf{0.71} & 0.01 & 0.70 & \textbf{0.10} & 2.25 & 0.05 \\
        & NVI & \textbf{0.14} & 0.01 & 0.14 & \textbf{1.19} & 9.50 & 0.12 \\
        & NID & \textbf{0.61} & 0.12 & 0.49 & \textbf{0.77} & 7.25 & 0.11 \\
        \midrule
        SIMCE & Binder & \textbf{0.19} & 0.06 & 0.12 & \textbf{9.50} & 14.32 & 0.66 \\
        & VI & \textbf{0.45} & 0.01 & 0.44 & \textbf{0.63} & 23.55 & 0.03 \\
        & NVI & \textbf{0.28} & 0.06 & 0.21 & \textbf{35.40} & 187.36 & 0.19 \\
        & NID & \textbf{0.44} & 0.10 & 0.34 & \textbf{22.07} & 133.07 & 0.17 \\
        \bottomrule
    \end{tabular}
\end{table}

Using these three sets, we first compare our implementation of the SALSO
algorithm to the algorithm of \citet{rastelli.friel.2018} as implemented in the
\texttt{MinimiseEPL} function of their \texttt{GreedyEPL} package, which is archived on CRAN.
\citet{rastelli.friel.2018} recommend using a single run of their stochastic
algorithm so, for the sake of comparison, we also limit SALSO to a single run.
For SALSO, we try at most 10 zealous updates, use a 50\% probability of
sequential allocation, and do not constrain the number of clusters. We use the
default settings for the \texttt{MinimiseEPL} function in the \texttt{GreedyEPL}
package. The results for the four loss functions supported by the
\texttt{GreedyEPL} package are shown in Table \ref{tab:simstudy1}. The ``A $<$
B'' column shows the proportion of times that the Monte Carlo estimate of the
posterior expected loss for the estimate reported by the SALSO algorithm was
lower than that obtained by the R\&F algorithm, while the ``B $<$ A'' column
shows the proportion of times that R\&F is better than SALSO. To aid comparison,
the difference between these two proportions is also noted. Note that the
proportion of ties is $1 - (\text{``A $<$ B''} + \text{``B $<$ A''})$. The mean
run time for each method is also recorded, as well as the ratio of the times.
From Table \ref{tab:simstudy1}, it is clear that the SALSO algorithm is
outperforming the algorithm proposed by \citet{rastelli.friel.2018} both in
terms of quality of the answer and in terms of the run time. The SALSO algorithm
yields an answer better than the R\&F algorithm much more than the R\&F
algorithm yields an answer better than the SALSO algorithm. The SALSO algorithm
also obtains an answer much more quickly on average.

\begin{table}[t]
    \footnotesize
	\centering
	\caption{\footnotesize Mean number of clusters and the run time of: (A) the
	SALSO algorithm with any number of clusters and (B) the SALSO algorithm
	constrained to yield an estimate with no more clusters than the maximum number
	of clusters among the samples. ``Binder(2)'' and ``VI(0.5)'' are losses with
	unequal weights.  The constrained optimization and the unequal weights are both
	able to successfully control the number of clusters and lead to faster
	computations.
		\newline
		\newline \textbf{A}: SALSO (10, 0.5), 1 run, unconstrained
		\newline \textbf{B}: SALSO (10, 0.5), 1 run, constrained \label{tab:simstudy2}}
	\begin{tabular}{llrrrrr}
		\toprule
		& & \multicolumn{2}{c}{\# of Clusters} & \multicolumn{3}{c}{Run Time} \\
		\cmidrule(lr){3-4}
		\cmidrule(lr){5-7}
		Dataset & Loss & A & B & A & B & Ratio \\
		\midrule
		PM10 & Binder & 6.36 & \textbf{6.00} & 0.01 & \textbf{0.01} & 1.12 \\
		& Binder(2) & 3.75 & \textbf{3.75} & 0.01 & \textbf{0.01} & 1.12 \\
		& omARI  & 4.81 & \textbf{4.69} & 0.02 & \textbf{0.02} & 1.19 \\
		& VI & 3.92 & \textbf{3.82} & 0.01 & \textbf{0.01} & 1.10 \\
		& VI(0.5) & 10.22 & \textbf{6.35} & 0.01 & \textbf{0.01} & 1.31 \\
		& NVI & 19.31 & \textbf{8.25} & 0.05 & \textbf{0.03} & 1.67 \\
		& ID & 6.35 & \textbf{6.33} & 0.03 & \textbf{0.03} & 1.13 \\
		& NID & 9.95 & \textbf{7.71} & 0.05 & \textbf{0.03} & 1.44 \\
		\midrule
		Gaskins & Binder & 30.59 & \textbf{16.34} & 0.26 & \textbf{0.10} & 2.55 \\
		& Binder(2) & 7.17 & \textbf{7.16} & 0.10 & \textbf{0.08} & 1.32 \\
		& omARI  & 16.76 & \textbf{14.65} & 0.46 & \textbf{0.32} & 1.43 \\
		& VI & 4.46 & \textbf{4.45} & 0.10 & \textbf{0.08} & 1.37 \\
		& VI(0.5) & 25.61 & \textbf{15.74} & 0.32 & \textbf{0.13} & 2.38 \\
		& NVI & 27.14 & \textbf{15.41} & 1.19 & \textbf{0.44} & 2.70 \\
		& ID & 13.21 & \textbf{12.96} & 0.58 & \textbf{0.48} & 1.21 \\
		& NID & 15.02 & \textbf{14.58} & 0.77 & \textbf{0.51} & 1.51 \\
		\midrule
		SIMCE & Binder & 261.11 & \textbf{10.98} & 9.50 & \textbf{0.32} & 29.61 \\
		& Binder(2) & 38.43 & \textbf{7.79} & 2.09 & \textbf{0.37} & 5.58 \\
		& omARI  & 71.11 & \textbf{9.63} & 8.61 & \textbf{1.12} & 7.69 \\
		& VI & 2.93 & \textbf{2.93} & 0.63 & \textbf{0.19} & 3.40 \\
		& VI(0.5) & 3.76 & \textbf{3.55} & 1.14 & \textbf{0.23} & 4.87 \\
		& NVI & 253.15 & \textbf{10.26} & 35.40 & \textbf{2.32} & 15.28 \\
		& ID & 44.58 & \textbf{9.68} & 10.36 & \textbf{1.83} & 5.66 \\
		& NID & 92.85 & \textbf{10.66} & 22.07 & \textbf{2.40} & 9.21 \\
		\bottomrule
	\end{tabular}
\end{table}

The next part of the study is meant to show the ability of the SALSO algorithm
and the chosen loss function to control the number of clusters. In the SALSO
algorithm, the maximum number of clusters can easily be set, which has important
implications for the interpretability of the resulting clustering and can also
influence the RAM and CPU time needed for the optimization algorithm. The
default in the \texttt{salso} package is to constrain the optimization by the
maximum number of clusters observed among the supplied posterior clusterings. We
considered the SALSO algorithm with default settings (10 zealous updates, 0.5
probability of sequential allocation) having both constrained (by the maximum
observed) and unconstrained number of clusters.  The results are shown in Table
\ref{tab:simstudy2}.  The constraint is clearly successful in limiting the
number of clusters.  Notice that especially for the SIMCE set, the number of
clusters is huge, leading to solutions that are hard to interpret and take
substantially more CPU time and, although not shown here, more RAM. Because the
constrained algorithm yields more practical results, we will use the constrained
algorithm for the remainder of the study.

Another important point from Table \ref{tab:simstudy2} is the success of the
generalized Binder and the generalized VI in controlling the number of clusters.
In addition to the four loss functions supported by the \texttt{GreedyEPL}
package, Table \ref{tab:simstudy2} also considers the omARI loss, the Binder
loss with $a = 2, b = 1$ (labeled ``Binder(2)''), and our generalized variation
of information with $a=0.5, b=1$ (labeled ``VI(0.5)''). The ``Binder(2)'' sets
the penalty for incorrectly separating items that should be clustered together
to be twice as much as the penalty for incorrectly clustering items that should
be separate. Similarly, the ``VI(0.5)'' loss represents the generalized VI loss
in which the penalty for incorrectly separating items is half that of
incorrectly clustering items.  The loss functions with unequal weights are
discussed in Section \ref{sec:generalized}. Note that ``Binder(2)'' yields
estimates with fewer mean number of clusters than the regular ``Binder''
estimates for both the unconstrained and constrained settings. The ``VI(0.5)''
clustering estimates have more mean number of clusters than the regular ``VI''
estimates for both settings. This shows that the modifications proposed in
Section \ref{sec:generalized} do indeed have the desired control over the number
of clusters.

\begin{table}[t]
    \footnotesize
	\centering
	\caption{\footnotesize The quality of: (A) the default SALSO
	algorithm with 4 runs, (B) the default SALSO algorithm with 1 run, and (C) our
	implementation of the \citet{rastelli.friel.2018} algorithm with multiple runs
	run as long as (A). Note that (C) is the SALSO algorithm without zealous
	updates or sequential allocation.
		\newline
		\newline \textbf{A}: SALSO (10, 0.5), 4 runs
		\newline \textbf{B}: SALSO (10, 0.5), 1 run
		\newline \textbf{C}: SALSO (0, 0.0), timed to 4 runs of SALSO (10, 0.5)\label{tab:simstudy3}}
	\begin{tabular}{llrrrrrr}
		\toprule
		& & \multicolumn{3}{c}{Quality} & \multicolumn{3}{c}{Quality} \\
		\cmidrule(lr){3-5}
		\cmidrule(lr){6-8}
		Dataset & Loss & A $<$ B & B $<$ A & Diff. & A $<$ C & C $<$ A & Diff. \\
		\midrule
		PM10 & Binder & \textbf{0.07} & 0.00 & 0.07 & \textbf{0.01} & 0.01 & 0.00 \\
		& Binder(2) & \textbf{0.01} & 0.00 & 0.01 & \textbf{0.01} & 0.00 & 0.01 \\
		& omARI & \textbf{0.05} & 0.00 & 0.04 & \textbf{0.02} & 0.01 & 0.01 \\
		& VI & \textbf{0.07} & 0.00 & 0.06 & \textbf{0.07} & 0.01 & 0.06 \\
		& VI(0.5) & \textbf{0.24} & 0.03 & 0.21 & \textbf{0.10} & 0.06 & 0.04 \\
		& NVI & \textbf{0.32} & 0.07 & 0.26 & \textbf{0.21} & 0.14 & 0.07 \\
		& ID & \textbf{0.26} & 0.04 & 0.22 & \textbf{0.12} & 0.06 & 0.06 \\
		& NID & \textbf{0.26} & 0.04 & 0.22 & \textbf{0.14} & 0.09 & 0.04 \\
		\midrule
		Gaskins & Binder & \textbf{0.75} & 0.19 & 0.56 & 0.44 & \textbf{0.46} & -0.03 \\
		& Binder(2) & \textbf{0.01} & 0.00 & 0.01 & \textbf{0.00} & 0.00 & 0.00 \\
		& omARI & \textbf{0.29} & 0.05 & 0.24 & 0.11 & \textbf{0.13} & -0.02 \\
		& VI & \textbf{0.13} & 0.01 & 0.13 & \textbf{0.13} & 0.02 & 0.12 \\
		& VI(0.5) & \textbf{0.62} & 0.14 & 0.48 & 0.34 & \textbf{0.34} & -0.01 \\
		& NVI & \textbf{0.58} & 0.13 & 0.45 & \textbf{0.33} & 0.31 & 0.02 \\
		& ID & \textbf{0.54} & 0.07 & 0.47 & \textbf{0.33} & 0.12 & 0.20 \\
		& NID & \textbf{0.49} & 0.07 & 0.42 &\textbf{0.24} & 0.16 & 0.08 \\
		\midrule
		SIMCE & Binder & \textbf{0.43} & 0.08 & 0.35 & \textbf{0.25} & \textbf{0.25} & 0.00 \\
		& Binder(2) &\textbf{0.20} & 0.05 & 0.15 & \textbf{0.15} & 0.12 & 0.03 \\
		& omARI & \textbf{0.32} & 0.09 & 0.22 & \textbf{0.26} & 0.16 & 0.09 \\
		& VI & \textbf{0.06} & 0.01 & 0.05 & \textbf{0.24} & 0.02 & 0.23 \\
		& VI(0.5) & \textbf{0.13} & 0.02 & 0.11 & \textbf{0.12} & 0.03 & 0.09 \\
		& NVI & \textbf{0.38} & 0.08 & 0.31 & \textbf{0.30} & 0.19 & 0.11 \\
		& ID & \textbf{0.38} & 0.09 & 0.29 & \textbf{0.26} & 0.16 & 0.10 \\
		& NID & \textbf{0.39} & 0.12 & 0.28 & \textbf{0.29} & 0.21 & 0.09 \\
		\bottomrule
	\end{tabular}
\end{table}

The next part of the study compares four runs of the SALSO algorithm with
default settings to a single run of the SALSO algorithm with default settings.
Table \ref{tab:simstudy3} shows these results in the comparison between (A) and
(B). Four runs of the SALSO algorithm obtains a better answer on average than
the single run in every case. In contrast to the recommendation of
\citet{rastelli.friel.2018}, we find that multiple runs are indeed beneficial.
Since the SALSO algorithm is embarrassingly parallel, these four runs can be
performed in essentially the same amount of time as a single run when using a
computer with four or more cores. Therefore, we recommend that multiple runs of
the SALSO algorithm be performed in order to obtain a better clustering
estimate.

When comparing with the implementation of \citet{rastelli.friel.2018} as shown
in Table \ref{tab:simstudy1}, the SALSO algorithm obtained a better answer in a
faster time than the R\&F algorithm on average. However, as noted in Section
\ref{sec:algorithms}, the \citet{rastelli.friel.2018} algorithm as implemented
in \texttt{GreedyEPL} does not control well the number of clusters and, further,
there could be differences in the efficiency of the implementations. We
therefore compare SALSO against the our implementation of the
\citet{rastelli.friel.2018} algorithm, using the same constraint on the number
of clusters for both methods. Specifically, we compare the SALSO algorithm with
the default settings (10 zealous updates, 0.5 probability of sequential
allocation) to the SALSO algorithm with 0 zealous updates and 0 probability of
sequential allocation, where the number of clusters is no more than that
observed among the supplied clusterings. Of course, for a fixed number of
iterations, SALSO with zealous updates can do no worse on average than SALSO
without zealous updates on average, but zealous updates add CPU cost.  It is
interesting to compare these two algorithms for a fixed time budget.  To this
end, both algorithms run for a fixed amount of time, which is roughly the time
that it takes for four runs of the SALSO algorithm with default settings. These
results are shown in Table \ref{tab:simstudy3} in the comparison between (A) and
(C). In the vast majority of cases, it appears that the SALSO algorithm with
default settings is obtaining a better answer on average than our implementation
of the R\&F algorithm, even though our implementation of the R\&F algorithm can
perform more runs in the fixed CPU budget.

We also attempted to investigate whether the benefits of the SALSO algorithm are
due largely to zealous updates, sequential allocation, or a combination of both.
Results suggest that the zealous updates are more important, but that sequential
allocation is also generally beneficial. Based on these results, we recommend
using the SALSO algorithm with zealous updates and the probability of sequential
allocation set to around 0.5, because there is not a clear winner between the
random allocation and sequential allocation for algorithm initialization. Doing
each about half the time would seem to provide opportunities to better explore
the partition space.

\section{Conclusion}
\label{sec:conclusion}

This paper addresses the problem of point estimation of a partition based on
samples from the posterior partition distribution.  We view the major
contributions of this paper as threefold.  First, we propose the generalized
variation of information (GVI) loss which allows for differential weights on the
two classification mistakes.  We prove that the GVI is a quasimetric and show
that it can effectively control the number of clusters.  Likewise, the
\citet{binder.1978} loss function, as originally proposed, allows for
differential weights and we note that the literature seems to have forgotten the
advantage of differential weights in controlling the number of clusters.  We
suspect this realization may revitalize Binder loss in the statistical
community. The second major contribution is the SALSO algorithm, which is a
novel greedy search algorithm over the space of partitions to minimize the Monte
Carlo estimate of the posterior expected loss.  One key aspect of the algorithm
is the so-called zealous updates.  The algorithm is amenable for any loss
function and can benefit from caching.  In the case of Binder loss and the GVI,
additional computational shortcuts allow for even more computational efficiency,
allowing for optimization for large values of $n$. The last major contribution
is a parallel, computationally-efficient implementation of the SALSO algorithm,
along with the new and many existing loss functions, in the form of the
\texttt{salso} package on CRAN.

\section{Appendix}
\label{sec:appendix}

\subsection{Proof of General Form of Binder Loss}

Here we prove the equivalence between the Binder loss in (\ref{binder}) --- as
it was originally stated by \cite{binder.1978} --- and our expression in terms
of the contingency table in (\ref{generalizedBinderNinvariant}).  As described
in Section \ref{sec:salso}, this lends itself to computationally efficient
optimization of the posterior expected loss. Note that $a, b > 0$ are the costs
of the misclassfication mistakes discussed in Section \ref{sec:generalized}.  We
use the notation in the contingency table of Table \ref{tab:contingency}.
\begin{align*}
	L&_{\text{binder}}(\textbf{c}, \hat{\textbf{c}}) = \sum_{i<j} \left(a\cdot\mathbb{I}\{c_i = c_j\}\mathbb{I}\{\hat{c}_i \neq \hat{c}_j\} + b\cdot\mathbb{I}\{c_i \neq c_j\}\mathbb{I}\{\hat{c}_i = \hat{c}_j\}\right)\\
    &= \sum_{i<j} \left(a\cdot\mathbb{I}\{c_i = c_j\}(1-\mathbb{I}\{\hat{c}_i = \hat{c}_j\}) + b\cdot(1-\mathbb{I}\{c_i = c_j\})\mathbb{I}\{\hat{c}_i = \hat{c}_j\}\right)\\
	&=a\sum_{i<j}\mathbb{I}\{c_i = c_j\} + b\sum_{i<j}\mathbb{I}\{\hat{c}_i = \hat{c}_j\} - (a+b)\sum_{i<j}\mathbb{I}\{c_i = c_j\}\mathbb{I}\{\hat{c}_i = \hat{c}_j\}\\
	&=a\sum_{i=1}^{k} {n_{i\cdot}\choose 2} + b\sum_{j=1}^{\hat{k}} {n_{\cdot j}\choose 2} - (a+b)\sum_{i=1}^{k} \sum_{j=1}^{\hat{k}} {n_{ij}\choose 2}\\
	&=	\frac{1}{2} \left( a \sum_{i=1}^{k} n_{i\cdot}^2 + b\sum_{j=1}^{\hat{k}}n_{\cdot j}^2 - (a+b)\sum_{i=1}^{k} \sum_{j=1}^{\hat{k}}n_{ij}^2
		- a\sum_{i=1}^{k}n_{i\cdot} - b\sum_{j=1}^{\hat{k}}n_{\cdot j} + (a+b)\sum_{i=1}^{k} \sum_{j=1}^{\hat{k}}n_{ij}\right)\\
	&= \frac{1}{2} \left( a\sum_{S \in \rho} |S|^2 + b\sum_{T \in \hat{\rho}} |T|^2 - (a+b)\sum_{S \in \rho} \sum_{T \in \hat{\rho}} |S \cap T |^2\right).
\end{align*}
\noindent Thus, the $n$-invariant version of the generalized Binder loss is:
\begin{equation*}
	L_{\text{n'binder}}(\textbf{c}, \hat{\textbf{c}}) = \frac{2}{n^2} L_{\text{binder}}(\textbf{c}, \hat{\textbf{c}})   = a \sum_{S \in \rho} \left( \frac{|S|}{n}\right)^2 + b \sum_{T \in \hat{\rho}} \left( \frac{|T|}{n}\right)^2 - (a+b) \sum_{S \in \rho} \sum_{T \in \hat{\rho}} \left( \frac{|S \cap T |}{n} \right)^2.
\end{equation*}

\subsection{Proofs of Quasimetric Properties}

Here we prove Theorems \ref{gbinder_quasimetric} and \ref{gvi_metric}, which
state that the generalized Binder loss and our GVI are quasimetrics. For Binder
loss, it is helpful to recall the definition in (\ref{binder}).  Likewise, for
the GVI, recall the identities in Section \ref{subsec:vi} and the definition in
(\ref{gvi_def}). In order to satisfy the quasimetric property, a distance
measure must have the following two properties: i.\ the identity of
indiscernibles and ii.\ the triangle inequality.

A loss function $L$ satisfies the identity of indiscernibles when $L(\rho,
\hat{\rho}) = 0 \iff \rho = \hat{\rho}$. \citet[see page 879]{meila.2007} states
that $ \rho = \hat{\rho} \iff I(\rho, \hat{\rho}) = H(\rho) = H(\hat{\rho})$.
Therefore the identity of indiscernibles holds, since the GVI can be written in
the following form:
\begin{align*}\label{gvi_pos_def}
    L_{\text{GVI}}(\rho, \hat{\rho}) &= bH(\rho) + aH(\hat\rho) - (a+b)I(\rho, \hat\rho)\\
    &=b\left\lbrace H(\rho) - I(\rho,\hat{\rho}) \right\rbrace + a\left\lbrace H(\hat{\rho}) - I(\rho,\hat{\rho}) \right\rbrace.
\end{align*}

The identity of indiscernibles is straightforward for the generalized Binder
loss as defined in (\ref{binder}). It is clear that if two clusterings are
equivalent, there will be no disagreements for any $ i,j $ pair, and thus 0 loss
will be incurred. Conversely, if the two clusterings are not equivalent, there
will be a disagreement for at least one $i,j$ pair, and a strictly positive loss
of either $a$ or $b$ will be incurred.

The triangle inequality requires that, for any three partitions $ \rho_1,
\rho_2, \rho_3 $ in the space of all possible partitions, $ L(\rho_1, \rho_2) +
L(\rho_2, \rho_3) \geq L(\rho_1, \rho_3)$. Note that this is equivalent to
proving that $ L(\rho_1, \rho_2) + L(\rho_2, \rho_3) - L(\rho_1, \rho_3) \geq
0$. Consider the following proof for the GVI:
\begin{align*}
	L_{\text{GVI}}&(\rho_1, \rho_2) + L_{\text{GVI}}(\rho_2, \rho_3) - L_{\text{GVI}}(\rho_1, \rho_3) \\
	&= \begin{aligned}[t]bH(\rho_1) &+ aH(\rho_2) - (a+b)I(\rho_1, \rho_2) + bH(\rho_2) + aH(\rho_3) - (a+b)I(\rho_2,\rho_3) \\
		&- bH(\rho_1) - aH(\rho_3) + (a+b)I(\rho_1,\rho_3)\end{aligned}	\\
	&= (a+b)[H(\rho_2) - I(\rho_1,\rho_2) - I(\rho_2,\rho_3) + I(\rho_1,\rho_3)]\\
	&= \begin{aligned}[t] (a+b)[H(\rho_2) &- (H(\rho_1) + H(\rho_2) - H(\rho_1,\rho_2)) - (H(\rho_2) + H(\rho_3) - H(\rho_2,\rho_3))\\
		&+ (H(\rho_1) + H(\rho_3) - H(\rho_1,\rho_3))]\end{aligned}\\
	&= (a+b)[ ( H(\rho_1,\rho_2) - H(\rho_2) ) + ( H(\rho_2,\rho_3) - H(\rho_3) ) - ( H(\rho_1,\rho_3) - H(\rho_3) )]\\
	&= (a+b)[H(\rho_1\mid \rho_2) + H(\rho_2\mid \rho_3) - H(\rho_1\mid \rho_3)]\\
	&\geq 0
\end{align*}
The final line follows because $ a,b > 0$ by definition and $[H(\rho_1\mid
\rho_2) + H(\rho_2\mid \rho_3) - H(\rho_1\mid \rho_3)] \geq 0 $ is established
by \citet[see proof of Property 1, (35)-(37)]{meila.2007}. Therefore the product
is greater than or equal to $0$, so the triangle inequality holds for the GVI.

Now, consider the following proof for the generalized Binder loss. Let the
clusterings \textbf{x}, \textbf{y}, and \textbf{z} correspond to the equivalent
partitions $ \rho_1 $, $ \rho_2 $, and $ \rho_3 $. For convenience, the notation
$ \mathbb{X}_{i=j} $ is used in place of the indicator function $
\mathbb{I}\{x_i = x_j\} $.
\begin{align*}
	&L_{\text{binder}}(\textbf{x}, \textbf{y}) + L_{\text{binder}}(\textbf{y}, \textbf{z}) -
	L_{\text{binder}}(\textbf{x}, \textbf{z}) \\
	&= \sum_{i<j} \left( \left( a\mathbb{X}_{i=j}\mathbb{Y}_{i \neq j} + b\mathbb{X}_{i \neq j}\mathbb{Y}_{i = j}\right) +
	\left( a\mathbb{Y}_{i=j}\mathbb{Z}_{i \neq j} + b\mathbb{Y}_{i \neq j}\mathbb{Z}_{i = j}\right) -
	\left( a\mathbb{X}_{i=j}\mathbb{Z}_{i \neq j} + b\mathbb{X}_{i \neq j}\mathbb{Z}_{i = j}\right) \right)\\
	&= a\sum_{i<j} \left( \mathbb{X}_{i=j}\mathbb{Y}_{i \neq j} + \mathbb{Y}_{i = j}\mathbb{Z}_{i \neq j} - \mathbb{X}_{i=j}\mathbb{Z}_{i \neq j} \right) +
	b\sum_{i<j}\left( \mathbb{X}_{i \neq j}\mathbb{Y}_{i = j} + \mathbb{Y}_{i \neq j}\mathbb{Z}_{i = j} - \mathbb{X}_{i \neq j}\mathbb{Z}_{i = j} \right)\\
	 &\geq 0
\end{align*}
The final line holds by the following logic. Since $ a > 0$ and $b > 0$ by
definition, we simply need to show that both sums are non-negative, since this
makes their positively weighted sum also non-negative. The only way that the
first sum could be negative is if there exists some pair $i$, $j$ such that $
\mathbb{X}_{i=j}\mathbb{Z}_{i \neq j} $ can evaluate to one while $
\mathbb{X}_{i=j}\mathbb{Y}_{i \neq j}$ and $\mathbb{Y}_{i = j}\mathbb{Z}_{i \neq
j} $ both evaluate to zero. If, for a specific $i,j$ pair, $x_i=x_j$ and
$z_i\neq z_j$, then $ \mathbb{X}_{i=j}\mathbb{Z}_{i \neq j} = 1 $. However, for
both of the other terms to evaluate to 0 (and thus yielding -1 for that
summand), we must have that $ x_i = x_j$, $y_i \neq y_j$, $y_i = y_j$, and $z_i
\neq z_j$. It is clear that $ y_i = y_j $ and $ y_i \neq y_j $ are mutually
exclusive, and therefore the first sum can never be negative. Similar logic can
be used to show that the second sum can never be negative. Therefore, the
triangle inequality holds for the generalized Binder loss.

\bibliographystyle{asa}
\bibliography{references}

\end{document}